\documentclass[12pt]{article}
\usepackage{amsfonts}

\usepackage{graphicx}
\usepackage{amsmath}
\usepackage{float}

\textwidth=6.5in \hoffset=-.75in \voffset=-1in \numberwithin{equation}{section}
\numberwithin{figure}{section}

\begin{document}

\begin{titlepage}
\bigskip \begin{flushright}
\end{flushright}
\vspace{1cm}
\begin{center}
{\Large \bf {Cosmological Solutions in Five Dimensional Minimal Supergravity}}\\
\end{center}
\vspace{1cm}
\begin{center}
 A. M.
Ghezelbash{ \footnote{ E-Mail: masoud.ghezelbash@usask.ca}}
\\
Department of Physics and Engineering Physics, \\ University of Saskatchewan, 
Saskatoon, Saskatchewan S7N 5E2, Canada\\
\vspace{1cm}
\end{center}
\begin{abstract}

We construct new non-stationary cosmological solutions to five-dimensional minimal supergravity that don't have 
any tri-holomorphic U(1) isometries.
Our new solutions, in part, contain some of the previously constructed
solutions to the minimal supergravity. 
The  c-function of solutions shows monotonic increasing/decreasing behaviour in time,  in agreement with the expected behaviour of c-function in spacetimes with positive cosmological constant.   

\end{abstract}
\bigskip
\end{titlepage}\onecolumn

\bigskip

\section{Introduction}

In the strong coupling limit of brane systems, many horizonless
three-charge brane configurations undergo a geometric transition and become
smooth horizonless geometries with black hole or black ring charges. These charges come completely from fluxes wrapping on non-trivial cycles.
The three-charge black hole (ring) systems are dual to the states of
corresponding CFTs: in favour of the idea that non-fundamental-black hole
(ring) systems effectively arise as a result of many horizonless
configurations \cite{B1}. 
In the heart of eleven-dimensional three-charged supergravity solutions, there  is a four-dimensional hyper-K\"{a}hler metric (which is equivalent to a metric
with self-dual curvature in four dimensions) that guarantees the solutions preserve some supersymmetries \cite{B3}. The
five-dimensional sub-space-time of the eleven-dimensional three-charged metric together with Maxwell field make the bosonic sector of five-dimensional minimal supergravity equivalent to the Einstein-Maxwell-Chern-Simons Theory.
The Einstein-Maxwell-(Dilaton-(Axion)) or -(Chern-Simons) theories in different dimensionalities have been extensively explored from many different directions. The black hole solutions have been considered in \cite{EY1} as well as solitonic and gravitational instantons, dyonic and pp-wave solutions in \cite{EY2}, supergravity solutions in \cite{EY3}, brane worlds and cosmology in \cite{EY4}, NUT and Bolt solutions, Liouville potential, rotating solutions and string theory extensions of Einstein-Maxwell fields in \cite{EY5}.
In five-dimensions, unlike the four dimensions that the only horizon topology
is 2-sphere, we can have different more interesting horizon topologies such
as black holes with horizon topology of 3-sphere \cite{Myers1}, black rings
with horizon topology of 2-sphere $\times $ circle \cite{Em1}, black
saturn: a spherical black hole\ surrounded by a black ring \cite{El1}, black
lens which the horizon geometry is a Lens space $L(p,q)$ \cite{Ch1}. All
allowed horizon topologies have been classified in \cite{Ca1}.
In \cite{B4, Gah}, the authors consider hyper-K\"{a}hler Atiyah-Hitchin and Einstein-hyper-K\"{a}hler triaxial Bianchi type IX  base
spaces to construct five-dimensional
supergravity solutions that only have rotational $U(1)$
isometries. The complete solutions are regular around the critical surface of
base spaces. 
The solutions in \cite{Gah} are quite remarkable because  
Einstein-hyper-K\"{a}hler Bianchi type IX geometry (that includes hyper-K\"{a}hler Atiyah-Hitchin as a special case) 
doesn't have any tri-holomorphic $U(1)$ isometry.
Hence the solutions could be used to study the interesting
physical processes such as, merger
of two Breckenridge-Myers-Peet-Vafa black holes \cite{Br1} or the geometric
transition of a three-charge supertube of arbitrary shape; that don't respect any tri-holomorphic $U(1)$ symmetry. 
We should emphasize that, in general, constructing solutions with non-tri-holomorphic $U(1)$ isometries is a rather complicated, 
tedious and challenging task. To our knowledge, for classical black holes and black rings, only two solutions exist \cite{B11}.

In this paper, we use Bianchi type IX space as the base space to
construct five-dimensional cosmological solutions to minimal supergravity with positive cosmological constant. The solutions enjoy generic non-tri-holomorphic $U(1)$ isometries. The idea behind this paper is the first step to search for and construct black hole solutions in the presence of positive cosmological constant on Bianchi type IX space. 

In fact, in \cite{Kasto}, the authors constructed multi-black hole solutions of Einstein-Maxwell theory in spacetimes with positive cosmological constant. The solutions describe an arbitrary number of charged black holes that are in motion with respect to each other. Moreover, the five and higher dimensional black hole solutions with positive cosmological constant were found in \cite{Londo} and \cite{Jap}. Specially in \cite{Jap}, the Eguchi-Hanson based black hole solutions are in a contracting phase derived by the cosmological constant, hence the solutions can describe coalescence of black holes in asymptotically de Sitter (dS) spacetimes. 
The Eguchi-Hanson space (as well as Atiyah-Hitchin space) is a special case of Bianchi type IX Einstein-K\"{a}hler space (see appendix C); hence this supports our idea to search for black hole solutions based on Bianchi type IX space, in spacetimes with positive cosmological constant.   

The outline of this paper is as follows. In section \ref{sec:BIX},
we give a brief review of self-dual Bianchi type IX space and five-dimensional minimal supergravity with cosmological constant. In section \ref{sec:MSS}, 
we present the class of cosmological non-stationary supergravity solutions over Bianchi type IX space and discuss the 
asymptotics of the solutions as well as the behaviour of $c$-function for the solutions. We conclude in section \ref{sec:CON} with a summary of our solutions and possible future research directions as well as three appendices.


\section{The Bianchi Type IX Space and Minimal Supergravity}
\label{sec:BIX}

The Bianchi type IX metric 
is locally given by the
following metric with an $SU(2)$ or $SO(3)$ isometry group \cite{GM} 
\begin{equation}
ds^{2}=e^{2\{A(\zeta )+B(\zeta )+C(\zeta )\}}d\zeta
^{2}+e^{2A(\zeta )}\sigma _{1}^{2}+e^{2B(\zeta )}\sigma _{2}^{2}+e^{2C(\zeta
)}\sigma _{3}^{2},  \label{BIXG}
\end{equation}%
where $\sigma _{i\text{ }}$'s are Maurer-Cartan one-forms. 
Integrating the Einstein equations (appendix A) as well as self-duality of the curvature imply
\begin{eqnarray}
\frac{dA}{d\zeta }&=&\frac{1}{2}\{e^{2B}+e^{2C}-e^{2A}\}-\alpha _{1}e^{B+C},\label{Ap} \\ 
\frac{dB}{d\zeta }&=&\frac{1}{2}\{e^{2C}+e^{2A}-e^{2B}\}-\alpha _{2}e^{A+C}, \label{Bp}\\ 
\frac{dC}{d\zeta }&=&\frac{1}{2}\{e^{2A}+e^{2B}-e^{2C}\}-\alpha _{3}e^{A+B},\label{Cp}%
\end{eqnarray}%
where $\alpha _{i},i=1,2,3$ are integration constants obeying 
$\alpha _{i}\alpha _{j}=\varepsilon _{ijk}\alpha _{k}
$.

The first set of solutions corresponds to 
$(\alpha _{1},\alpha _{2},\alpha _{3})=(1,1,1)$ that yields Atiyah-Hitchin metric (appendix B). The Atiyah-Hitchin metric (and its ambi-polar extension) was considered extensively in \cite{B4} for construction of supergravity/Einstein-Maxwell-Chern-Simons solutions. 
The second set of solutions corresponds to $(\alpha _{1},\alpha _{2},\alpha _{3})=(0,0,0)$.
Changing the coordinate from $\zeta$ to $r$ given by (\ref{rzeta})    (appendix C), we find the metric of triaxial Bianchi type IX space as
\begin{equation}
ds^{2}=\frac{dr^{2}}{\sqrt{F(r)}}+\frac{r^{2}}{4}%
\sqrt{F(r)}\left( \frac{\sigma _{1}^{2}}{1-\frac{a_{1}^{4}}{r^{4}}}+\frac{%
\sigma _{2}^{2}}{1-\frac{a_{2}^{4}}{r^{4}}}+\frac{\sigma _{3}^{2}}{1-\frac{%
a_{3}^{4}}{r^{4}}}\right),   \label{BIX}
\end{equation}%
where
\begin{equation}
F(r)=\prod_{i=1}^{3}(1-\frac{a_{i}^{4}}{r^{4}}),  \label{FF}
\end{equation}%
and $a_{1},a_{2}$ and $a_{3}$ are three parameters that without loss of
generality, we choose them such that $0=a_{1}\leq a_{2}=2kc\leq $ $a_{3}=2c$%
. We note coordinate $r$ must be greater or equal to $a_3$. \ Here $0\leq k\leq 1$ is the square root of 
modulus of different types of Jacobian
elliptic functions (appendix C) and $c>0.$ If $k>1,$ all we need is just to interchange
the $2$ and $3$ directions. 
To embed the Bianchi space into five-dimensional minimal supergravity with cosmological constant, we take the metric ansatz as
\begin{equation}
ds^{2}=-H(t,\zeta)^{-2}dt^{2}+H(t,\zeta)ds_{4}^2,
\label{ds5}
\end{equation}%
and the only non-vanishing component of gauge field as
\begin{equation}
A_t=\frac{\eta\sqrt{3}}{2}\frac{1}{H(t,\zeta)},
\label{gauge5}
\end{equation}%
where $\eta=\pm 1$. 
The five-dimensional minimal supergravity with a positive cosmological constant is described by the action 
\begin{equation}
S=\frac{1}{16\pi}\int d^5x\, \sqrt{-g}(R-4\Lambda -F_{\mu\nu}F^{\mu\nu}-\frac{2}{3\sqrt{3}}\epsilon^{\mu\nu\rho\eta\xi}F_{\mu\nu}F_{\rho\eta}A_\xi),
\end{equation}
where $R$ and  $F_{\mu\nu}$ are five-dimensional Ricci scalar and Maxwell field.
The Einstein and Maxwell equations are
\begin{eqnarray}
R_{\mu\nu}-(\frac{1}{2}R-2\Lambda) g_{\mu\nu}&=&2(F_{\mu\lambda}F_{\nu}^{\lambda}-\frac{1}{4}g_{\mu\nu}F^2),\label{EEQ} \\
F^{\mu\nu}_{;\nu}&=&\frac{2}{3\sqrt{3}}\epsilon^{\mu\nu\rho\eta\xi}F_{\nu\rho}F_{\eta\xi},\label{GEQ}
\end{eqnarray}
respectively.

\section{Minimal Supergravity Solutions With Cosmological Constant}
\label{sec:MSS}

We find the solutions to supergravity equations (\ref{EEQ}) and (\ref{GEQ}) assuming the five-dimensional metric, the embedded four-dimensional space and gauge field are given by (\ref{ds5}), (\ref{BIXG}) and (\ref{gauge5}) respectively. All gravitational equations of motion are satisfied provided $H(t,\zeta)$ is a solution to the
differential equations,
\begin{eqnarray}
\frac{3\frac{\partial ^2 H(t,\zeta)}{\partial \zeta ^2}}{e ^{2(A(\zeta)+B(\zeta)+C(\zeta))}}-3(\frac{\partial H(t,\zeta)}{\partial t})^2H^2(t,\zeta)+4\Lambda H^2(t,\zeta)&=&0, \\
3(\frac{\partial H(t,\zeta)}{\partial t})^2+3\frac{\partial ^2 H(t,\zeta)}{\partial t^2}H(t,\zeta)-4\Lambda&=&0, \\
\frac{\partial ^2 H(t,\zeta)}{\partial \zeta \partial t}&=&0.
\label{eqforH}
\end{eqnarray}
\begin{figure}[tbp]
\centering           
\begin{minipage}[c]{.5\textwidth}
        \centering
        \includegraphics[width=\textwidth]{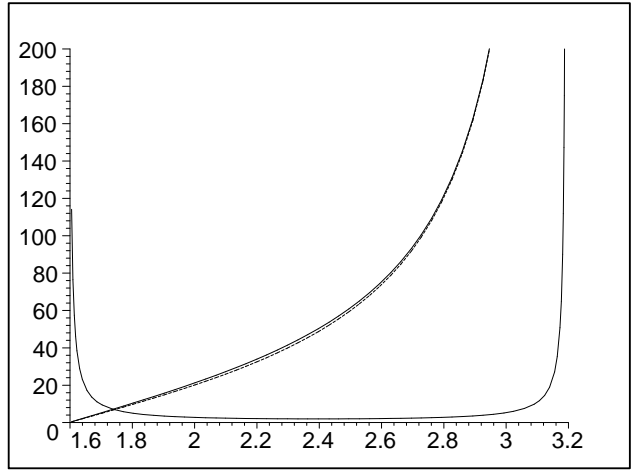}
    \end{minipage}
\caption{
The metric functions $50e^{2A(\zeta)}$ (dashed curve), $50e^{2B(\zeta)}$ (solid curve slightly above the dashed one) and $e^{2C(\zeta)}$ (almost symmetric solid curve) versus $\zeta$ where we set $c=1$ and $k=1/2$.
}
\label{ABCmetricfunctions}
\end{figure}

Taking $H(t,\zeta)=2\epsilon \sqrt{\frac{\Lambda}{3}}t+m\zeta+\gamma $
where $\epsilon=\pm 1$ and $m,\gamma$ are two constants of integration, all the equations of motion are satisfied, hence we obtain the supergravity metric as
\begin{eqnarray}
ds^{2}=&-&\frac{dt^2}{\Big(2\epsilon \sqrt{\frac{\Lambda}{3}}t+m\zeta+\gamma\Big)^2}\nonumber \\
&+&\Big(2\epsilon \sqrt{\frac{\Lambda}{3}}t+m\zeta+\gamma\Big)\Big(e^{2\{A(\zeta )+B(\zeta )+C(\zeta )\}}d\zeta
^{2}+e^{2A(\zeta )}\sigma _{1}^{2}+e^{2B(\zeta )}\sigma _{2}^{2}+e^{2C(\zeta
)}\sigma _{3}^{2}\Big).  \label{5DBIXG} \nonumber \\
&&
\end{eqnarray}
The metric functions $A(\zeta)$, $B(\zeta)$ and $C(\zeta)$
are given by equations (\ref{A1}), (\ref{A2}) and (\ref{A3}) respectively and their exponentials plotted in figure (\ref{ABCmetricfunctions}), where we set $c=1$ and $k=1/2$.

We would prefer to find the analytic solutions to supergravity equations of motion to embed triaxial Bianchi type IX space (\ref{BIX}) in (\ref{ds5}). However, we find too unlikely to get the analytic solutions despite the metric (\ref{BIX}) looks simpler in structure than (\ref{BIXG}).

\begin{figure}[tbp]
\centering           
\begin{minipage}[c]{.5\textwidth}
        \centering
        \includegraphics[width=\textwidth]{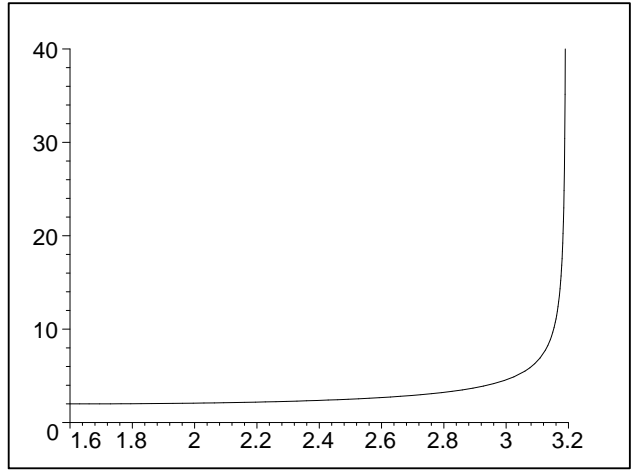}
    \end{minipage}
\caption{
The radial coordinate $r \geq 2c$ in (\ref{BIX}) as a function of coordinate $\zeta$ in (\ref{BIXG})where we set $c=1$ and $k=1/2$. 
}
\label{rzetafig}
\end{figure}

We note in figure (\ref{ABCmetricfunctions}), the coordinate $\zeta$ in (\ref{BIXG}) varies as $\zeta_0/2 \leq \zeta \leq \zeta_0$ covers the range of $0 \leq r < \infty$ where 
$\mathfrak{sn}(\zeta=\zeta_0 ,1/4)=0$ and $\zeta_0=3.192484+{\cal O}(10^{-7})$ (See figure (\ref{rzetafig})). In general for any $c$ and $0<k<1$, the coordinate $\zeta$ should be chosen as $\zeta_{m,c,k}/2 \leq \zeta \leq \zeta_{m,c,k}$ where  $\zeta_{m,c,k}$ is the $m$-th positive root of $\mathfrak{sn}(c^2\zeta ,k^2)$. 

For both $\epsilon=\pm 1$, we choose $m,\gamma >0$ (The mass of solutions is proportional to parameter $m$ that appears in metric (\ref{5DBIXG})). The choice makes the solution (\ref{5DBIXG}) with $\epsilon=+1$ regular everywhere for $t \geq 0$. Where $\epsilon=-1$, the solutions are still regular everywhere for $t \leq 0$. 

The Ricci scalar of solution (\ref{5DBIXG}) is given by 
\begin{equation}
R=\frac{20G(t,\zeta)-27/2m^2e^{-2(A(\zeta)+B(\zeta)+C(\zeta))}}{H^3(t,\zeta)},
\label{Riccis}
\end{equation}
where 
\begin{eqnarray}
G(t,\zeta)&=&27
\Lambda\gamma^2m\zeta+9\Lambda m^3\zeta^3+36\Lambda^2t^2m\zeta+27\Lambda\gamma m^2\zeta^2+36\Lambda^2t^2\gamma+9\Lambda\gamma^3\nonumber\\
&+&\epsilon(8\sqrt{3}\Lambda^{5/2}t^3+18\sqrt{3}\Lambda^{3/2}t m^2\zeta^2+18\sqrt{3}\Lambda^{3/2}t\gamma^2+36\sqrt{3}\Lambda^{3/2}t\gamma m\zeta),
\end{eqnarray}
and
\begin{equation}
H(t,\zeta)=2\sqrt{3}\epsilon\Lambda^{1/2}t+3\gamma+3m\zeta.
\end{equation}
For the highest possible value of $\zeta$ where $\zeta \rightarrow \zeta_0$, the metric (\ref{5DBIXG}) reduces to 
\begin{equation}
ds^2=-dT^2+e^{2\epsilon\sqrt{\Lambda/3}T}ds^2_{\mathbb{R}^{4}},
\label{asymzeta0}
\end{equation} 
where $ds^2_{\mathbb{R}^{4}}$ is given by (\ref{s3METRIC}) and $T=\frac{\ln(2\epsilon\sqrt{\Lambda/3}t+m\zeta_0+\gamma)}{2\epsilon\sqrt{\Lambda/3}}$.
The equal time hypersurfaces in (\ref{asymzeta0}) are flat $\mathbb{R}^{4}$. These hypersurfaces have an infinitely large size at $T \rightarrow -\infty$ (where $\epsilon=-1$), which decreases to a minimum value of $1$ as $T \rightarrow 0$. Hence the five-dimensional spacetime (\ref{asymzeta0}) with $\epsilon=-1$ shows a collapsing patch of five-dimensional dS spacetime.  This contracting patch of solutions (\ref{5DBIXG}) at future infinity implies that black hole solutions based on Bianchi type IX space can describe the coalescence of black holes in asymptotically dS spacetimes.                                          
Indeed, 
the Ricci scalar (\ref{Riccis}) of (\ref{asymzeta0}) is equal to $\frac{20\Lambda}{3}$, that is exactly the Ricci scalar of five-dimensional dS spacetime \cite{Leb}. T
The flat hypersurfaces $\mathbb{R}^{4}$ have the smallest size at $T=0$ (where $\epsilon=+1$), which increases to an infinitely large size as $T \rightarrow \infty$. Hence the spacetime (\ref{asymzeta0}) with $\epsilon=+1$ shows an expanding patch of five-dimensional dS spacetime. 
So in the limit of $\zeta \rightarrow \zeta_0$,
the spacteime (\ref{5DBIXG}) asymptotically reduces to the expanding/collapsing patches of five-dimensional dS spacetime \cite{Leb}.

In the other extreme case where $\zeta=\zeta_0/2+\hat\zeta$ with $\hat \zeta \rightarrow 0^+$,  
the solution (\ref{5DBIXG}) transforms to
\begin{equation}
ds^2=-d{\cal T}^2+e^{2\epsilon\sqrt{\Lambda/3}{\cal T}}\{ d\hat\zeta ^2+\sigma_1^2+\sigma_2^2+\frac{1}{\lambda^2}\sigma_3^2\},
\label{asymzeta02}
\end{equation}
where ${\cal T}=\frac{\ln\{\lambda(2\epsilon\sqrt{\Lambda/3}t+m\zeta_0/2+\gamma)\}}{2\epsilon\sqrt{\Lambda/3}}$ and we denote $e^{2A(\zeta_0/2)}=e^{2B(\zeta_0/2)}=e^{-2C(\zeta_0/2)}$ by $\lambda \rightarrow 0^+$.  
The metric (\ref{asymzeta02}) clearly shows a bolt structure at $\hat \zeta =0$ in which both Ricci scalar and Kretschman invariant diverge as $\frac{1}{\lambda}$ and $\frac{1}{\lambda ^6}$, respectively.
\begin{figure}[tbp]
\centering           
\begin{minipage}[c]{.5\textwidth}
        \centering
        \includegraphics[width=\textwidth]{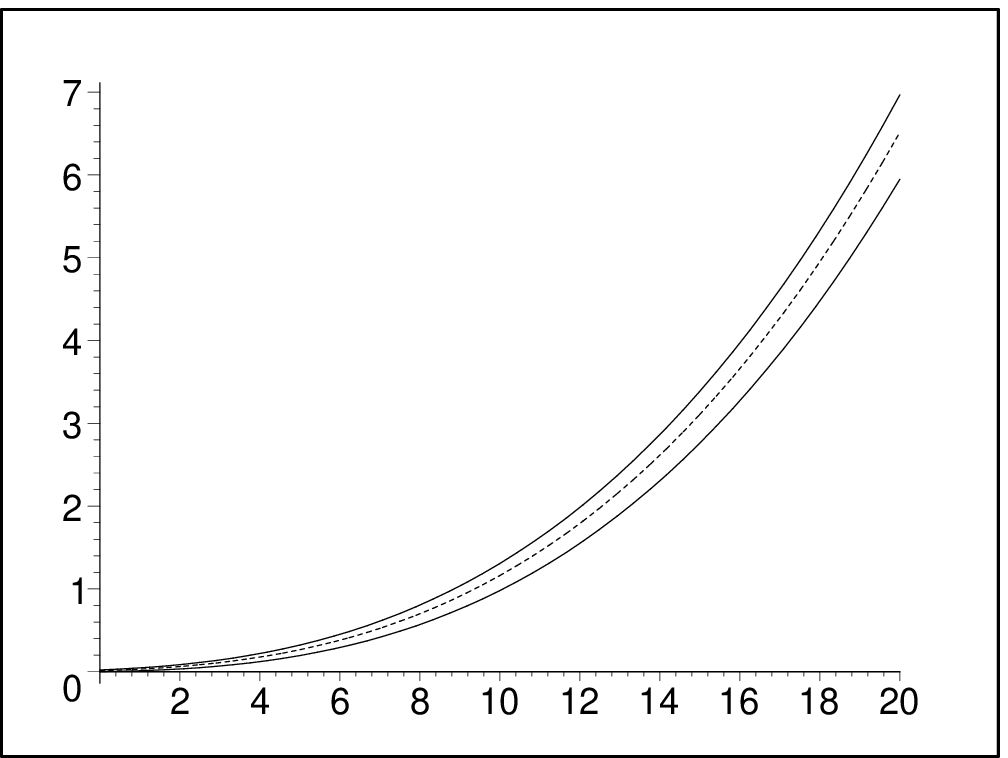}
    \end{minipage}
\caption{
The $c$-function$/10^3$ versus time (for $\epsilon=+1$) at three different $\zeta$-fixed slices, where we set $c=1$ and $k=1/2$. The up and down solid curves are $c$-functions at $\zeta=3.19$ and $1.60$ slices, respectively. The dashed curve shows the $c$-function for slice at $\zeta=2.50$.
}
\label{cfunctions}
\end{figure}

Finally, we consider the geometry of solutions (\ref{5DBIXG}) where $\epsilon=+1$ at future infinity. The metric in the limit of large values of $t$ approaches
\begin{equation}
ds^2=-d{\bf T}^2+\mu e^{\mu {\bf T}}ds^2_{B},
\label{5Dtlarge}
\end{equation} 
where ${\bf T}=\frac{\ln t}{\mu}$, $\mu=2\sqrt{\frac{\Lambda}{3}}$ and $ds^2_B$ is given by (\ref{BIXG}).
The equal time hypersurfaces of (\ref{5Dtlarge}) are Bianchi spaces.
The metric (\ref{5Dtlarge}) shows the big bang patch which covers half the dS spacetime from a big bang at past horizon to the Bianchi hypersurfaces at future infinity. The other half of dS spacetime is covered by the big crunch patch that includes the Bianchi hypersurfaces at past infinity to a big crunch at future horizon. This contracting patch of solutions (\ref{5DBIXG}) at future infinity implies that black hole solutions based on Bianchi type IX space can describe the coalescence of black holes in asymptotically dS spacetimes.

\begin{figure}[tbp]
\centering           
\begin{minipage}[c]{.5\textwidth}
        \centering
        \includegraphics[width=\textwidth]{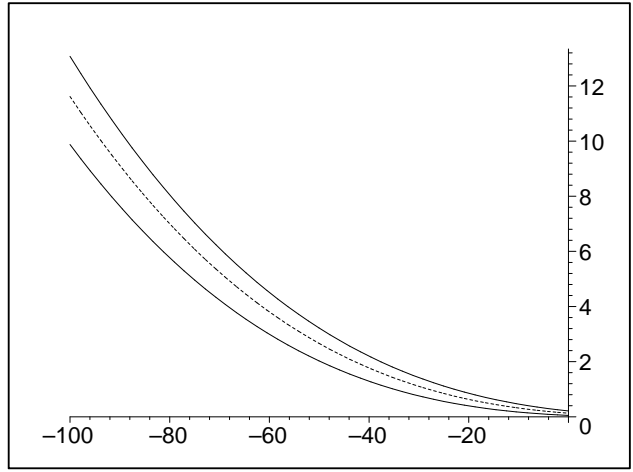}
    \end{minipage}
\caption{
The $c$-function$/10^5$ versus time for $\epsilon=-1$, at three different $\zeta$-fixed slices, where we set $c=1$ and $k=1/2$. The up and down solid curves are $c$-functions at $\zeta=3.19$ and $1.60$ slices, respectively. The dashed curve shows the $c$-function for slice at $\zeta=2.50$.
}
\label{cfunctionsM}
\end{figure}

As it is well known, there is a natural correspondence between
phenomena occurring near the boundary (or in the deep interior) of 
asymptotically dS (or AdS) spacetime and UV (IR) physics in the dual CFT. 
Hence, any solutions in 
asymptotically (locally) dS spacetims lead to interpretation in terms of
renormalization group flows and associated generalized dS $c$-theorem.
In any contracting patch of dS spacetimes, the
renormalization group flows toward the infrared and in any expanding
patch, it flows toward the ultraviolet. 
The $c$-function for representation of the dS metric with a wide
variety of boundary geometries involving direct products of flat space, the
sphere and hyperbolic space was studied in \cite{Leb1}. For our five-dimensional exact cosmological solution (\ref{5DBIXG}), the $c$-function is 
$ c=\left( G_{\mu \nu }n^{\mu }n^{\nu }\right) ^{-\frac{3}{2}}
$ 
where $n^{\mu }$\ is the unit vector along time direction. The $c$-function should show an increasing (decreasing) behaviour versus time for any expanding (contracting) patch of the spacetime. 
The explicit expression for $c$-function is given by 
\begin{eqnarray}
c^{-2/3}(t,\zeta)&=&\frac{3}{2}\frac{(\frac{\partial H(t,\zeta)}{\partial t})^2}{H^2(t,\zeta)}\nonumber\\
&+&\frac{3e^{-2(A(\zeta)+B(\zeta)+C(\zeta))}}{4H^5(t,\zeta)}\big((\frac{\partial H(t,\zeta)}{\partial \zeta})^2-2H(t,\zeta)\frac{\partial ^2 H(t,\zeta)}{\partial \zeta^2}  \big)
\end{eqnarray}
which shows the $c$-function depends on time as well as $\zeta$. We consider three different values for $\zeta$ and evaluate the corresponding $c$-function as  function of time. First, we consider $\zeta=\zeta _0 /2 \simeq 1.60$ that is the smallest value of $\zeta$. Second, we set $\zeta=2.50$ and finally we consider $\zeta=\zeta _0 \simeq 3.19$.
In figure (\ref{cfunctions}), we plot the $c$%
-functions for these three different fixed values of $\zeta$ where we set $\epsilon=+1$. The plots show explicitly the expansion of $\zeta$-fixed slices, in perfect agreement with what we concluded after equation   
(\ref{asymzeta0}). Moreover, setting $\epsilon=-1$ yields the decreasing $c$-functions of figure (\ref{cfunctionsM}), again in agreement with our previous conclusions.


\section{Conclusions}
\label{sec:CON}

In this paper, we construct an exact cosmological class of non-stationary solutions to five dimensional minimal supergravity, based on Bianchi type IX Einstein-hyperk\"{a}hler space. The Bianchi type IX Einstein-hyperk\"{a}hler space doesn't have any tri-holomorphic $U(1)$ isometries, hence the
solutions could be used to study the physical processes that don't respect
any tri-holomorphic abelian symmetries. The solutions are regular everywhere except on the location of bolt in four-dimensional Bianchi type IX base space. 
We note the $c$-function for our solutions depends on cosmological time as well as radial coordinate of Bianchi type IX base space. For any fixed value of coordinate $\zeta$, the $c$-function has monotonically increasing (or decreasing) behaviour with time depending on $\epsilon=+1$ (or $\epsilon=-1$). The behaviour of $c$-function is in perfect agreement with asymptotic reduction of solutions to expanding (or collapsing) patches of five-dimensional dS spacetimes.
We also notice the geometry of solutions at future infinity contains a contracting patch. This implies that black hole solutions based on Bianchi type IX space can describe the coalescence of black holes in asymptotically dS spacetimes.
We leave the black hole solutions based on Bianchi type IX space as well as the thermodynamics of solutions and application of dS/CFT correspondence to the solutions presented in this paper for a future article.

\bigskip
\bigskip
\noindent{\Large Acknowledgments}

This work was supported by the Natural Sciences and Engineering Research
Council of Canada.

\bigskip

{\Large Appendix A}

\bigskip

The Maurer-Cartan one-forms in (\ref{BIXG}) are given respectively by 
\begin{eqnarray}
\sigma _{1}&=&d\psi +\cos \theta d\phi, \label{s1}\\ 
\sigma _{2}&=&-\sin \psi d\theta +\cos \psi \sin \theta d\phi,\label{S2} \\ 
\sigma _{3}&=&\cos \psi d\theta +\sin \psi \sin \theta d\phi, \label{S3}%
\end{eqnarray}%
and satisfy 
\begin{equation}
d\sigma _{i}=\frac{1}{2}\varepsilon _{ijk}\sigma _{j}\wedge \sigma _{k}.
\label{dsigma}
\end{equation}%
We note that the metric on the $\mathbb{R}^{4}$ (with a radial coordinate $R$
and Euler angles ($\theta ,\phi ,\psi $) on an $S^{3}$) could be written in
terms of Maurer-Cartan one-forms via%
\begin{equation}
ds^{2}=dR^{2}+\frac{R^{2}}{4}(\sigma _{1}^{2}+\sigma _{2}^{2}+\sigma
_{3}^{2}),  \label{s3METRIC}
\end{equation}%
with $\sigma _{2}^{2}+\sigma _{3}^{2}$\ is the standard metric of $S^{2}$\
with unit radius; $4(\sigma _{1}^{2}+\sigma _{2}^{2}+\sigma _{3}^{2})$\
gives the same for $S^{3}.$ The Bianchi type IX metric (\ref{BIXG}) satisfies Einstein equations provided%
\begin{eqnarray}
\frac{d^{2}A}{d\zeta ^{2}}&=&\frac{1}{2}\{e^{4A}-(e^{2B}-e^{2C})^{2}\}, \label{App}\\ 
\frac{d^{2}B}{d\zeta ^{2}}&=&\frac{1}{2}\{e^{4B}-(e^{2C}-e^{2A})^{2}\}, \label{Bpp}\\ 
\frac{d^{2}C}{d\zeta ^{2}}&=&\frac{1}{2}\{e^{4C}-(e^{2A}-e^{2B})^{2}\},\label{Cpp}%
\end{eqnarray}%
\newline
as well as%
\begin{equation}
\frac{dA}{d\zeta }\frac{dB}{d\zeta }+\frac{dB}{d\zeta }\frac{dC}{d\zeta }+%
\frac{dC}{d\zeta }\frac{dA}{d\zeta }=\frac{1}{2}%
\{e^{2(A+B)}+e^{2(B+C)}+e^{2(C+A)}\}-\frac{1}{4}\{e^{4A}+e^{4B}+e^{4C}\}.
\label{Ein2}
\end{equation}%
Integrating equations (\ref{App}), (\ref{Bpp}), (\ref{Cpp}) and (\ref{Ein2}%
), we get equations (\ref{Ap}),(\ref{Bp}) and (\ref{Cp}) where $\alpha_i$'s are three integration constants. 

\bigskip

{\Large Appendix B}

\bigskip

The first set of solutions to equations (\ref{Ap}), (\ref{Bp}) and (\ref{Cp}) corresponds to $(\alpha _{1},\alpha _{2},\alpha _{3})=(1,1,1)$. The solutions describe the Atiyah-Hitchin metric \cite{Hana} in the general form of (\ref{BIXG}). The metric functions are
\begin{eqnarray}
e^{2A(\zeta )}&=&\frac{2}{\pi }\frac{\vartheta _{2}\vartheta _{3}^{^{\prime
}}\vartheta _{4}^{^{\prime }}}{\vartheta _{2}^{^{\prime }}\vartheta
_{3}\vartheta _{4}},\label{e2A} \\ 
e^{2B(\zeta )}&=&\frac{2}{\pi }\frac{\vartheta _{2}^{^{\prime }}\vartheta
_{3}\vartheta _{4}^{^{\prime }}}{\vartheta _{2}\vartheta _{3}^{^{\prime
}}\vartheta _{4}},\label{e2B} \\ 
e^{2C(\zeta )}&=&\frac{2}{\pi }\frac{\vartheta _{2}^{^{\prime }}\vartheta
_{3}^{^{\prime }}\vartheta _{4}}{\vartheta _{2}\vartheta _{3}\vartheta
_{4}^{^{\prime }}},\label{e2C}%
\end{eqnarray}%
where the $\vartheta $'s are Jacobi Theta functions with complex modulus $%
i\zeta $. The Jacobi Theta functions $\vartheta $ are given by $
\vartheta _{i}(\tau )=\vartheta _{i}(0\left| \tau \right.)
$, where we have used Jacobi-Erderlyi notation $\vartheta _{1}(\nu \left| \tau \right. )=\vartheta \left[ _{1}^{1}\right]
(\nu \left| \tau \right. ), \,
\vartheta _{2}(\nu \left| \tau \right. ) =\vartheta \left[ _{0}^{1}\right]
(\nu \left| \tau \right. ),  \, 
\vartheta _{3}(\nu \left| \tau \right. ) =\vartheta \left[ _{0}^{0}\right]
(\nu \left| \tau \right. ) $ and $
\vartheta _{4}(\nu \left| \tau \right. ) =\vartheta \left[ _{1}^{0}\right]
(\nu \left| \tau \right. ) $.

The Jacobi functions with characteristics 
$\vartheta \left[ _{b}^{a}\right] (\nu \left| \tau \right. )$ are defined by the following series
\begin{equation}
\vartheta \left[ _{b}^{a}\right] (\nu \left| \tau \right. )=\sum_{n\in
Z}e^{i\pi (n-\frac{a}{2})\{\tau (n-\frac{a}{2})+2(\nu -\frac{b}{2})\}},
\label{thy}
\end{equation}%
where $a$ and $b$ are two real numbers.
%

The other possible values for the integration constants $\alpha_1,\alpha_2$ and $\alpha_3$ are $(1,-1,-1)$, $(-1,1,-1)$ and $(-1,-1,1)$.  However, in all these three cases, a redefinition of metric functions change the solutions into Atiyah-Hitchin metric with metric functions (\ref{e2A}), (\ref{e2B}) and (\ref{e2C}).  

\bigskip

{\Large Appendix C}

\bigskip

The second set of solutions to equations (\ref{Ap}), (\ref{Bp}) and (\ref{Cp}) corresponds to $(\alpha _{1},\alpha _{2},\alpha _{3})=(0,0,0)$.  We find the differential
equations (\ref{Ap}), $(\ref{Bp})$ and (\ref{Cp}) can be solved exactly and the solutions are given by  \cite{Gah}
\begin{eqnarray}
A(\zeta)&=&\frac{1}{2}\ln \big ( {c^{2}
\frac{\mathfrak{cn}(c^2\zeta ,k^2)\mathfrak{dn}(c^2\zeta ,k^2)}{\mathfrak{sn}(-c^2\zeta ,k^2)}}
\big ), \label{A1} \\
B(\zeta)&=&\frac{1}{2}\ln \big ( {c^{2}
\frac{\mathfrak{cn}(c^2\zeta ,k^2)}{\mathfrak{dn}(c^2\zeta ,k^2)\mathfrak{sn}(-c^2\zeta ,k^2)}}
\big ),\label{A2} \\
C(\zeta)&=&\frac{1}{2}\ln \big ( {c^2
\frac{\mathfrak{dn}(c^2\zeta ,k^2)}{\mathfrak{cn}(c^2\zeta ,k^2)\mathfrak{sn}(-c^2\zeta ,k^2)}}
\big ),\label{A3} 
\end{eqnarray}
where $\mathfrak{sn}(z,k)$, $\mathfrak{cn}(z,k)$ and $\mathfrak{dn}(z,k)$; the standard Jacobian elliptic $SN$, $CN$ and $DN$ functions, are related
to $\mathfrak{am}(z,k)$; the Jacobian elliptic $AM$ function, by%
\begin{eqnarray}
\mathfrak{sn}(z,k)&=&\sin (\mathfrak{am}(z,k)),  \label{SN}\\
\mathfrak{cn}(z,k)&=&\cos (\mathfrak{am}(z,k)),  \label{CN}\\
\mathfrak{dn}(z,k)&=&\sqrt{1-k^2\mathfrak{sn}^2(z,k)}.  \label{DN}
\end{eqnarray}%
The Jacobian elliptic $AM$ function is the inverse of the trigonometric form
of the elliptic integral of the first kind; which means%
$\mathfrak{am}(\mathfrak{f}(\sin \phi ,k),k)=\phi
$, where $\mathfrak{f}(\varphi ,k)$; the elliptic integral of the first kind, is
defined by 
\begin{equation}
\mathfrak{f}(\varphi ,k)=\int_{0}^{\sin ^{-1}(\varphi )}\frac{d\theta }{%
\sqrt{1-k^{2}\sin ^{2}\theta }}.  \label{FG1}
\end{equation}
We change the 
coordinate $\zeta $ in the metric (\ref{BIXG})
to the coordinate $r$ by \cite{Gah}
\begin{equation}
r=\frac{2c}{\sqrt{\mathfrak{sn}(c^2\zeta ,k^2)}}.  \label{rzeta}
\end{equation}%
The metric then changes from (\ref{BIXG}) into (\ref{BIX}) with one metric function (\ref{FF}) only. The latter form of metric is more convenient to be considered in special cases, as we consider here.
In special case of $k=0$, where $a_1$ and $a_2$ coincide, we
get the metric%
\begin{equation}
ds^{2}=\frac{dr^{2}}{h(r)}+\frac{r^{2}}{4}h(r)\{d\theta ^{2}+\sin
^{2}\theta d\phi ^{2}\}+\frac{r^{2}}{4h(r)}(d\psi +\cos \theta d\phi )^{2},
\label{EHI}
\end{equation}%
which is the Eguchi-Hanson type I metric with $h(r)=(1-\frac{(2c)^{4}}{r^{4}}%
)^{1/2}$. 
The Eguchi-Hanson type I metric can also be written as \cite{EH} 
\begin{equation}
ds^{2}=\widetilde{f}^{2}(r)dr^{2}+\frac{r^{2}}{4}\widetilde{g}%
^{2}(r)\{d\theta ^{2}+\sin ^{2}\theta d\phi ^{2}\}+\frac{r^{2}}{4}(d\psi
+\cos \theta d\phi )^{2},  \label{EHIs}
\end{equation}%
where%
\begin{equation}
\begin{array}{c}
\widetilde{f}(r)=\frac{1}{2}(1+\frac{1}{\sqrt{1-\frac{a^{4}}{r^{4}}}}), \\ 
\widetilde{g}(r)=\sqrt{\frac{1}{2}(1+\sqrt{1-\frac{a^{4}}{r^{4}}})}.%
\end{array}
\label{EHIsmff}
\end{equation}%
In the other extreme case where $k=1$, $a_2$ is equal to $a_3$
and we obtain the Eguchi-Hanson type II metric%
\begin{equation}
ds^{2}=\frac{dr^{2}}{h^{2}(r)}+\frac{r^{2}}{4}h^{2}(r)\sigma _{1}^{2}+%
\frac{r^{2}}{4}(\sigma _{2}^{2}+\sigma _{3}^{2}),  \label{EHII2}
\end{equation}%
which is of the same form of well-known Eguchi-Hanson metric 
\begin{equation}
ds^{2} =\frac{r^{2}}{4g(r)}\left[ d\psi +\cos (\theta )d\phi \right]
^{2}+g(r)dr^{2}+\frac{r^{2}}{4}\left( d\theta ^{2}+\sin ^{2}(\theta )d\phi
^{2}\right),
\end{equation}%
by making the substitution $2c=a$ and $h(r)=\frac{1}{\sqrt{g(r)}}$ in (\ref{EHII2}). 
We note that only for special values of $k=0$ and $k=1,$ the metric (\ref{BIX}%
) admits a tri-holomorphic $U(1)$ isometry; hence could be put into
Gibbons-Hawking form. In both special cases of $k=0$ and $k=1$, the
five-dimensional supergravity solutions can be constructed simply by four
harmonic functions on the base space. The case with $k=1$ was considered
explicitly in \cite{To1} and \cite{coni}, where the authors constructed five-dimensional 
supersymmetric black ring solutions as well as eleven dimensional solutions on the four and six-dimensional hyper-K\"{a}hler Eguchi-Hanson
type II base spaces, respectively. Their solutions have the same two angular momentum
components and the asymptotic structure on time slices is locally Euclidean. 
The circle-direction of the black ring is along the
equator on a two-sphere bolt on the base space. The case with $k=0$ gives a
separable five-dimensional metric for Eguchi-Hanson type I manifold with a time
direction. The most general five-dimensional supergravity solutions with  parameter $k$ varies as $0< k \leq $ $1,$ were studied in \cite{Gah}.

\end{document}